\documentclass[12pt,preprint]{aastex}
\begin{document}

\title {Probing the Central Regions of 
Nearby Compact Elliptical Galaxies \altaffilmark{1}}

\author{T. J. Davidge }

\affil{Herzberg Institute of Astrophysics,
\\National Research Council of Canada, 5071 West Saanich Road,
\\Victoria, B.C. Canada V9E 2E7\\ {\it email: tim.davidge@nrc.ca}}

\author{Tracy L. Beck}

\affil{Gemini Observatory, 670 North A'ohoku Place,
\\Hilo, HI 96720-2700\\ {\it email: tbeck@gemini.edu}}

\author{Peter, J. McGregor}

\affil{Research School of Astronomy and Astrophysics, 
\\ Australian National University, Cotter Road, Weston Creek, 
\\ ACT 2611, Australia \\ {\it email: peter@mso.anu.edu.au}}

\altaffiltext{1}{Based on observations obtained at the
Gemini Observatory, which is operated by the Association of Universities
for Research in Astronomy, Inc., under a co-operative agreement with the
NSF on behalf of the Gemini partnership: the National Science Foundation
(United States), the Particle Physics and Astronomy Research Council
(United Kingdom), the National Research Council of Canada (Canada),
CONICYT (Chile), the Australian Research Council (Australia), CNPq (Brazil),
and CONICET (Argentina).}

\begin{abstract}

	$K-$band spectroscopic observations recorded with NIFS$+$ALTAIR 
on Gemini North are used to probe the central arcsec of the compact 
elliptical galaxies NGC 4486B, NGC 5846A, and M32. 
The angular resolution of these data is $\sim 0.1$ arcsec FWHM; this corresponds to 
a spatial scale of 12 parsecs in NGC 5846A, which is the most distant galaxy in the 
sample. Indices that probe the strengths of various 
atomic and molecular features are measured. The central stellar contents 
of NGC 4486B and NGC 5846A are similar, in the sense that they occupy 
the same regions of the (Ca I, $^{12}$CO), (Na I, $^{12}$CO) and ($^{13}$CO,$^{12}$CO) 
diagrams. The NGC 4486B and NGC 5846A observations depart from 
the sequence defined by solar neighborhood giants in the 
(Na I, $^{12}$CO) diagram, in a sense that is consistent with 
both galaxies having non-solar chemical mixtures. 
For comparison, the M32 data is consistent with a chemical enrichment 
history like that in the Galactic disk; M32 could not have formed from the stripping 
of a larger elliptical galaxy. The behaviour of the near-infrared 
line indices as a function of radius is also investigated. The stellar content in the 
central arcsec of M32 appears to be well mixed. However, the radial behaviour of the 
indices in NGC 4486B and NGC 5846A show complicated behaviour, with the gradients that are 
present at large radii breaking down or reversing within a few tenths of an arcsec of 
the nucleus. Based on the age gradients predicted from visible wavelength spectra, coupled 
with the radial behaviour of the $<FeI>$ and $^{12}$CO(2,0) indices, 
it is suggested that the nuclear regions of NGC 4486B and NGC 5846A harbour intermediate 
age populations. 

\end{abstract}


\section{INTRODUCTION}

	Compact elliptical (cE) galaxies are among the rarest type of dwarf galaxy. 
Of the three cEs known within $\sim 30$ Mpc (M32, NGC 4486B, and NGC 5846A), all 
are neighbors of much more massive galaxies, and all appear to be tidally truncated (e.g. 
King 1962; Rood 1965, King \& Kiser 1973). There are other high surface brightness compact 
galaxies in the local universe with young populations that are interacting with larger 
companions, such as NGC 1510, that may be cEs in distinct stages of their evolution. 
Despite the tantalizing hints that interactions with a much larger galaxy may play a 
key role in the formation of cEs, these objects do not appear in large numbers in nearby 
galaxy clusters, and many candidate cEs have turned out to be background ellipticals 
(e.g. Ziegler \& Bender 1998; Drinkwater \& Gregg 1998). A modest number of cEs 
are seen in more distant clusters, such as Abell 496 (Chilingarian et al. 2005) and 
Abell 1689 (Mieske et al. 2005).

	The structural properties of cEs are reminiscent of larger classical elliptical 
galaxies (e.g Wirth \& Gallagher 1984). Indeed, cEs fall along the low brightness end of 
the trend defined by classical elliptical galaxies on the M$_B$ -- $<\mu>_e$ and 
R$_e$ -- $<\mu>_e$ planes. Given their modest brightnesses and the nature of these 
trends, the structural properties of cEs are thus `extreme' when compared with more 
massive elliptical galaxies, in the sense of having comparatively high central surface 
brightnesses ($<\mu>_e \leq 19$) and small effective radii (log(R$_{e(kpc)}) \leq 0.2$). 

	Various cE formation mechanisms have been examined in the literature, and 
most of these involve the tidal stripping of a progenitor galaxy. 
Faber (1973) suggested that cEs may be the remnant cores of ellipticals that have 
had their outer regions removed. Nieto \& Prugniel (1987) propose a variation of this 
idea, in which the progenitors of cEs are a special type of low mass 
high central surface brightness elliptical that have 
been stripped of their outer envelopes, although candidate 
progenitors have not yet been discovered. Bekki et al. (2001) suggest that M32 is 
the remnant bulge of a `threshed' disk galaxy, and evidence for a residual 
disk has been found (Graham 2002; Choi, Guhathakurta, \& Johnston 2002). 
The Bekki et al. (2001) model has the merit of explaining the paucity of cEs, 
as they only form in a small region of parameter space. It should be noted that 
tidal stripping is not the only process that has been considered to form cEs, 
and Burkert (1994) describes how a high central surface brightness galaxy may 
form in close proximity to a much larger galaxy.

	The characterization of stellar populations within galaxies, especially in their 
central regions where supermassive black holes are often found 
and signatures of recent star formation may also be present, provide information 
that is essential for probing the origins and evolution of these objects. 
In the current paper, the radial distribution of the brightest evolved stars 
in the central arcsec of the three closest cEs -- M32, NGC 4486B, and NGC 5846A -- 
is investigated. The data are unique in that they were obtained with an 
integral field spectrograph at near-infrared wavelengths, and have an angular resolution 
approaching the diffraction limit of an 8 meter telescope.

	While cEs share common structural properties, there are 
substantial galaxy-to-galaxy differences in their stellar contents.
As one of the closest and brightest galaxies in the sky, M32 
has been the subject of numerous spectroscopic and photometric investigations. There is 
spectroscopic evidence of an intermediate age population both at 
visible (e.g. O'Connell 1980; Rose 1985; Bica, Alloin, \& Schmidt 1990; del Burgo 
et al. 2001) and near-infrared (Davidge 1990) wavelengths. Studies of the resolved 
stellar content of M32 also reveal signs of an intermediate age component, in that 
the most luminous AGB stars have photometric properties that are indicative of an 
age of a few Gyr (e.g. Freedman 1992; Elston \& Silva 1992; 
Davidge 2000; Davidge \& Jensen 2007). Still, there is some 
disagreement between the analysis of the resolved stellar content and 
predictions made from integrated light spectroscopy. 
Davidge \& Jensen (2007) find that the brightest AGB stars in M32 are 
uniformly mixed throughout the galaxy, and this is contrary to what might be expected 
based on the radial gradients in age and metallicity that were found by 
Rose et al. (2005) and Worthey (2004) in their investigation of visible wavelength spectra.

	The stellar content of NGC 4486B differs from 
that of M32 in that the strengths of absorption features in 
the visible light spectrum of NGC 4486B are consistent with those of an old, 
metal-rich system (Sanchez-Blazquez et al. 2006b, hereafter SGCG). In addition, 
the absorption features in the visible spectrum of NGC 4486B are indicative of a 
metallicity gradient with a slope that is similar to those in 
classical elliptical galaxies (Sanchez-Blazquez, Gorgas, \& Cardiel 
2006, hereafter SGC), whereas the metallicity gradient in M32 is much shallower (e.g. 
Worthey 2004; SGC). There is also evidence for a 
comparatively steep age gradient in NGC 4486B when compared with other 
ellipticals (SGC), opening the possibility that an intermediate age component 
like that in M32 might await discovery near the center of NGC 4486B. 

	NGC 4486B harbours a double nucleus 
(Lauer et al. 1996), and there is kinematic evidence for a central 
super-massive black hole (Kormendy et al. 1997). 
Adopting the black hole mass estimated by Kormendy et al. (1997), then 
the ratio of central black hole to total bulge mass in NGC 4486B is larger 
than what is seen in other elliptical galaxies (e.g. Magorrian et al. 1998). One possible 
explanation of such a result is that NGC 4486B may once have been a more massive 
galaxy, but that the outer regions were stripped away, presumably by interactions with the 
much larger NGC 4486. The amount of material that would have to be 
stripped from the NGC 4486B progenitor in such a scenario is substantial. 
If the progenitor of NGC 4486B had a central black hole mass that fell along the 
trend between black hole mass and spheroid (`bulge') mass that 
is defined by other systems, then the progenitor spheroid must have been 
between one and two orders of magnitude more massive than is presently seen.
Still, the mass of the central object in NGC 4486B is uncertain, and 
Soria et al. (2006a) argue for a black hole mass that 
is an order of magnitude lower than that computed by Kormendy et al. (1997). 

	NGC 5846A has not been as extensively studied as M32 or NGC 4486B, probably 
because it is more distant. Still, NGC 5846A shares some similarities 
with NGC 4486B. Like NGC 4486B, NGC 5846A is a companion to a much larger galaxy, 
which in turn is the largest member of a group that is dominated by early-type galaxies 
(Mahdavi, Trentham, \& Tully 2005). As in NGC 4486B, the absorption lines at visible 
wavelengths in the integrated spectrum of NGC 5846A have strengths that are 
indicative of an old, metal-rich system (SGCG), and there is a metallicity gradient 
that is similar to those in classical ellipticals (SGC). The age gradient 
in NGC 5846A is shallower than that in NGC 4486B, but is still typical of that seen 
in other ellipticals (SGC).

	The spectra that have been used previously to probe the stellar contents of M32, 
NGC 4486B, and NGC 5846A were typically recorded with ground-based telescopes during 
natural seeing conditions, and so have angular resolutions that are on the order of an 
arcsec. Observations with higher angular resolution are desireable to probe the central 
regions of these galaxies, where young or intermediate age populations 
might lurk. The use of adaptive optics (AO) systems on ground-based 
8 metre telescopes can produce a substantial gain in angular resolution over 
what can be achieved without wavefront correction. In the 
current study, AO-corrected spectra of M32, NGC 4486B, 
and NGC 5846A in the $2 - 2.5\mu$m wavelength region are used to examine the 
stellar content of the central regions of these galaxies. The integrated light 
near $2\mu$m in old and all but the youngest of intermediate age systems 
is dominated by first ascent giants (e.g. Frogel et al. 1978; Maraston 2005). Given that 
the photometric properties of giants are sensitive to age and metallicity then 
spectroscopic observations in the $2 - 2.5\mu$m interval are potentially useful for 
investigating population gradients. 

	Various properties of the three cEs are listed in Table 1. The quantities 
listed for each galaxy are the distance modulus (second column), the absolute 
$K-$band magnitude (third column), the radius at which the surface brightness reaches 
25 mag arcsec$^{-2}$ in $B$ (fourth column), and the central velocity dispersion measured 
in a 0.1 arcsec radius aperture (fifth column). For homogeneity, the 
distance moduli are those measured by Tonry et al. (2001) from 
surface brightness flucuations. The absolute 
brightnesses were measured from images in the 2MASS Large Galaxy Atlas (Jarrett et al. 
2003), while the isophotal radii are from De Vaucouleurs et al. (1991). The velocity 
dispersions were measured from the data discussed here (\S 4). 

	The spectra were recorded with NIFS on Gemini North. 
The wavefront incident to NIFS was corrected by the ALTAIR AO system, and the 
resulting signal at the NIFS image plane has an angular resolution of $\sim 0.1$ arcsec. 
The resulting spatial resolution of these data spans a broad range. In M32 the spatial 
resolution is 0.4 parsecs, whereas in NGC 4486B the 
spatial resolution is $\sim 8$ parsecs and in NGC 5846A it is $\sim 12$ parsecs.

	The paper is structured as follows. Details of the 
observations and the techniques used to reduce 
the data are discussed in \S 2. The angular resolution of the data and a preliminary 
galaxy-to-galaxy comparison of the $K-$band spectra are the subjects of 
\S 3. A quantitative investigation of the stellar content using line indices, 
including an investigation of radial trends in stellar content, 
is carried out in \S 4. A summary and discussion of the results follows in \S 5.

\section{OBSERVATIONS \& REDUCTIONS}

	The spectra of all three galaxies were recorded with NIFS$+$ALTAIR 
on Gemini North (GN). NIFS (McGregor et al. 2003), is an integral field 
spectrograph that covers the $0.9 - 2.5\mu$m wavelength interval. 
The science detector is a $2048 \times 2048$ HAWAII-2RG HgCdTe array. 
A contiguous $3 \times 3$ arcsec$^2$ area on the sky is 
sampled with 29 slitlets, each of which subtends a $0.1 \times 3$ arcsec$^2$ area; the 
direction perpendicular to the dispersion axis has an angular scale of 0.04 arcsec 
pixel$^{-1}$, and the spectral dispersion is 5300.

	ALTAIR (Herriot et al. 2000) is the facility AO system on GN. 
Prior to 2006, ALTAIR used natural guide stars (NGSs) exclusively
as reference beacons for AO correction, and the M32 data were recorded 
in this mode, with the bright semi-stellar galaxy nucleus serving as the NGS. 
This mode could not be employed for NGC 4486B and NGC 5846A as 
the nucleus of each galaxy is too faint and too diffuse to serve as a NGS. Moreover, 
both galaxies are at moderately high Galactic latitudes, 
and there are no stars of a suitable brightness that are close enough on 
the sky to serve as NGSs. Therefore, the observations of 
NGC 4486B and NGC 5846A were obtained using laser guide stars 
(LGSs), which is a capability that has recently been commissioned 
on GN (Boccas et al. 2006).

	When using artificial reference beacons it is necessary 
to monitor the tip/tilt (T/T) component of the incident wavefront 
with a natural guide source near the science target. While 
the nuclei of NGC 4486B and NGC 5846A are not suitable for use as NGSs 
for high-order AO correction, they can be used as T/T reference sources. This is 
possible since the inherently low angular resolution of the four element T/T wavefront 
sensor (WFS) in ALTAIR makes it more stable when guiding on faint, extended sources 
than the higher-order NGS WFS.

	The data were recorded over two observing seasons. 
The ALTAIR field lens, which moves the conjugatation 
point for AO correction from a fixed altitude above the summit of 
Mauna Kea to the primary mirror of the telescope, was deployed for these 
observations, as the use of this lens significantly reduces 
anisoplanicity. The M32 spectra were obtained on the night of October 23, 
2005 as part of the NIFS commissioning program. The observations of NGC 4486B and NGC 
5846A were recorded as part of programs GN2006B-SV-103 and GN2007A-DD-6. 
The spectra of NGC 4486B used in this paper were obtained on the night of February 
6, 2007, while the observations of NGC 5846A were recorded on June 1, 2007. 
Six 600 sec exposures were recorded of M32, while 
eight 85 sec exposures were recorded of NGC 4486B and 
fourteen 85 sec exposures were recorded of NGC 5846A.

	Each galaxy observation was paired with that 
of a blank sky field, which was recorded either immediately before or after 
the galaxy observation. At least one telluric standard star observation was made 
either immediately before or after the galaxy observations on a given night. 
In the case of NGC 4486B and NGC 5846A, the telluric standard was 
observed with the LGS AO system activated so that the corrected image quality could 
be assessed. A modest sample of bright spectroscopic/radial velocity standard stars 
were also observed as part of program GN2006B-SV-103. The telluric and spectroscopic 
standard stars are summarized in Table 2.

	The first step in the data reduction sequence was to subtract the sky spectra 
from the corresponding galaxy exposures, and the results 
were divided by a flat-field frame. Cosmic rays were identified using a routine
that identified pixels with intensities that greatly differed from those of the 
surroundings. The intensities of the pixels that were found to contain cosmic rays 
were set to the median of the surrounding pixels. The flat-fielded data were 
wavelength calibrated using observations of an Ar arc, which 
were recorded during each night. The wavelength-calibrated spectra 
for each galaxy were then summed, and the combined spectra were divided by the spectrum 
of the appropriate telluric standard star. With the exception of Br$\gamma$, the spectra 
of the telluric standard stars are largely free of photospheric absorption lines. The 
Br$\gamma$ line was removed from the telluric spectra using the 
interactive line-fitting and subtraction routine in the IRAF routine SPLOT.

\section{DATA CHARACTERIZATION}

\subsection{Assessing the Angular Resolution of the Observations}

	The angular resolution of these data is an important issue when considering 
the spectroscopic characteristics of the central regions of the target galaxies. 
To monitor the intrinsic seeing, the telluric standard stars that were observed as part of 
the NGC 4486B and NGC 5846A data sets were recorded with 
the LGS AO system activated. The PSFs obtained from 
the telluric star observations are centrally concentrated, with a FWHM $\sim 0.1$ arcsec. 
At 0.2 arcsec radius the PSF drops to $\sim 5\%$ of the peak value, while at 0.4 arcsec 
radius the PSF is only $\sim 1\%$ of the peak value.

	There is crosstalk between the spatial elements in 
integral field spectrographs because of (1) the size of the source being studied, 
which includes contributions not only from the intrinsic angular extent of the object, 
but also from seeing and the telescope optics, and (2) 
the optics within the spectrograph. The issue of slitlet-to-slitlet scatter 
was investigated using the light profiles of the telluric standard stars 
that were observed with ALTAIR activated. If the NIFS optics introduce 
significant slitlet-to-slitlet scattering then the FWHM of the stellar 
profile measured along the central slitlet will be narrower than the FWHM measured 
perpendicular to this direction. The light profile of the telluric standard stars 
measured between slitlets is not wider than that measured along the central slitlet, 
indicating that the AO-corrected PSF is the dominant contributor to 
crosstalk between NIFS slitlets for sources with FWHM = 0.1 arcsec.

	The nature of AO-corrected point spread functions (PSFs) 
depends on environmental conditions and the properties of the T/T source. 
The image quality of the galaxy observations can be assessed in a manner that 
accounts for these factors by comparing the AO-corrected light profiles of the 
galaxies with those obtained using the HST. To enable such a comparison, light 
profiles were constructed for each galaxy by integrating the two dimensional 
spectra between rest frame wavelengths $2.20$ and $2.27\mu$m, and the results are shown in 
Figure 1. All profiles plotted in this figure have been normalized to the peak value.

	Two of the galaxies in our sample have published light profiles based on 
HST data. The dashed line in the top panel of Figure 1 is the F475W light profile of 
NGC 4486B from Ferrarese et al. (2006), and there is remarkable agreement with 
the NIFS data. Thus, the angular resolution of the NGC 4486B NIFS data are 
comparable to that obtained with the HST at visible wavelengths. 
The dashed line in the lower panel of Figure 1 is the F222M light profile 
for M32 plotted in Figure 8 of Corbin, O'Neil, \& Rieke (2001). The NICMOS data 
have an angular resolution $\sim 0.2$ arcsec. The ALTAIR observations are significantly 
narrower than the HST data when $r < 0.4$ arcsec, indicating that the current data has a 
higher angular resolution than was acheived with NICMOS.

	NGC 5846A has the most compact light profile in our sample. 
Unfortunately, we are not aware of extant high angular resolution images of 
this galaxy. Still, given that the light profile has a width of $\pm 0.1$ arcsec, the 
NIFS observations of NGC 5846A likely have an angular resolution that is comparable 
to, or possibly even better than, that of the NGC 4486B data.

\subsection{The $2\mu$m Spectra: A First Look}

	The visible wavelength spectra of M32, NGC 4486B, and NGC 5846A 
show distinct differences, in the sense that M32 has deeper H$\beta$ and shallower 
Mgb absorption than the other galaxies. This in turn indicates that these 
galaxies have different stellar contents (e.g. SGCG). This being the 
case, then how do the spectra of these galaxies compare in the near-infrared, where the 
integrated light is dominated by stars that are very different from those that contribute 
significantly at visible wavelengths? The first overtone $^{12}$CO band heads are the most 
prominent features in the near-infrared spectra of old and intermediate age 
populations, and so are potentially prime diagnostics of stellar content. The depths 
of these features have been modelled previously, and they are predicted to be 
sensitive to population parameters (e.g. Figure 11 of Vazquez, Carigi, \& 
Gonzalez 2003).

	The $2\mu$m spectrum of the central arcsec of each galaxy, which is 
the approximate seeing disk during `typical' ground-based observing 
conditions, is shown in Figure 2. Also plotted are the spectra of the K5III star 
HR 4801 and the M0III star HR 4884. A low-order continuum has been divided out of each 
spectrum. The vertical axis in Figure 2 shows intensity in a linear scale, 
and the spectra have been shifted vertically to facilitate comparisons.

	The random and systematic motions of stars broaden absorption lines in the 
integrated spectra of galaxies, and spectra should be smoothed to a common kinematic 
broadening before making comparisons. To allow this to 
be done, a velocity dispersion and radial velocity were 
measured for each spectrum using the Tonry \& Davis (1979) cross-correlation routine. 
The velocity dispersions measured within 0.1 arcsec radius of the galaxy centers are 
listed in the last column of Table 1. The estimated uncertainties in these measurements 
are roughly $\pm 10$ km sec$^{-1}$. It should be emphasized that the entries in 
this Table are based on the widths of the absorption features averaged over a 0.1 
arcsec radius, and thus reflect contributions from both the random {\it and} rotational 
motions of stars in this region of each galaxy. The rotational component will contribute 
to the line widths, and smaller dispersions would be found had the spectra 
not been radially averaged, but instead restricted to the dynamical axes of the 
galaxy.

	The $\sigma_{0.1}$ entry for M32 in the last column of Table 1 is consistent with 
that measured by Joseph et al. (2001) from STIS spectra. The results in that paper 
predict a total line broadening of $\sim 130$ km sec$^{-1}$ 
in a 0.1 arcsec radius region centered on the galaxy nucleus. This 
agrees with the entry for this galaxy in Table 1, and so there is not a 
significant difference with the NIFS data. Spectra 
with moderately high angular resolution are also available for NGC 4486B. 
Kormendy et al. (1997) measure a velocity dispersion of $\sim 300$ km 
sec$^{-1}$ near the center of NGC 4486B, and this is 30 km sec$^{-1}$ lower than 
that found here. While the difference is larger than the estimated uncertainty in the 
NIFS measurements, the Kormendy et al. (1997) data have an angular resolution 
of $\sim 0.4$ arcsec, compared with $\sim 0.1$ arcsec for the NIFS data.

	The spectra plotted in Figure 2 have been Gaussian smoothed to match the velocity 
dispersion in the central regions of NGC 4486B, which has the highest velocity dispersion 
in the data (Table 1). The $^{12}$CO band heads in M32 are weaker than in the other 
galaxies, having a depth that is slightly shallower than in the K5III star. The depths 
of the Na I, Fe I, and Ca I features in M32 are also weaker than in the other galaxies, 
but are comparable to those in the K5III spectrum. 

	The spectra of NGC 4486B and NGC 5846A are similar, as might be 
expected based on the ages and metallicities of these galaxies 
inferred from visible wavelength spectra (e.g. SGCG). Still, 
the $^{12}$CO band heads are slightly deeper in NGC 5846A. 
This difference in line strength can be attributed -- 
at least in part -- to differences in the radial distributions of populations 
in the innermost regions of these galaxies (\S 4). 

\section{STELLAR CONTENTS}

\subsection{A Comparison of Stellar Contents Using Line Indices}

	Spectra were extracted in 0.1 arcsec wide annuli. The 0.1 arcsec 
width matches the angular resolution of the data while also
allowing the extracted spectra to have reasonably high signal-to-noise ratios. 
A velocity dispersion and a radial velocity were measured for each 
extracted spectrum. The velocity dispersion measures the line broadening caused 
by the random and systematic motions of stars in each annulus. 
Given that the extracted spectra sample annuli, as opposed to the kinematic 
axes, the velocity measurements are not suitable for a kinematic study; 
rather, they simply provide the empirical measure of line width 
that is needed to produce spectra with a common intrinsic resolution.

	To create spectra from which a homogeneous set of line strengths could 
be obtained, the extracted spectra were convolved with Gaussians so that the 
results had line broadenings that matched those in the 
central regions of NGC 4486B. The standard deviation 
of the Gaussian that was applied to each extracted spectrum was 
computed from the velocity dispersion measurements discussed above. 
The equivalent widths of Na I, Ca I, and $^{12}$CO (2,0) absorption features 
were measured in the broadened spectra using the 
continuum and line bandpasses defined by Ramirez et al. (1997). 

	Two other indices were measured. Silva, Kuntschner, \& Lyubenova (2007) define 
indices that sample Fe I transitions, and the results 
are averaged to produce the $<FeI>$ index. In addition, 
the $^{13}$CO(2,0) band head is seen in the spectra, and the equivalent 
width of this feature was measured using the bandpasses listed in Table 3. 
The $^{13}$CO(2,0) band head is located in the wings of the much 
stronger $^{12}$CO(4,2) feature, and the continuum passbands in Table 3 measure only 
a local pseudo-equivalent width. Hence, the $^{13}$CO(2,0) index defined here
is only a relative measure of the strength of this feature.

\subsubsection{The Impact of Kinematic Broadening}

	The impact of kinematic broadening on the line indices was assessed by convolving 
the standard star spectra with a Gaussian to simulate the 
velocity smearing near the center of NGC 4486B. Smoothing 
to match the kinematic properties of stars in the central regions of the cE galaxies has a 
substantial impact on the visual appearance of the stellar spectra. This is demonstrated 
in Figure 3, where the unsmoothed and smoothed spectra of the M0III giant HR 4884 
are compared. The absorption features in the smoothed 
spectrum are heavily blurred, and the complex of lines near the Na I and 
Ca I features each form a single broad trough in the smoothed spectrum.

	We gauge the impact of velocity broadening on line indices with the statistic 
C($\sigma$), which is the ratio of a line index measured from an unbroadened spectrum to 
that measured after applying a broadening function to simulate a velocity dispersion 
$\sigma$ (e.g. Davies, Sadler, \& Peletier 1993). The C($\sigma_{NGC4486B}$) values 
measured for the K1III star HR4793 and the M0III star HR 4884, where 
$\sigma_{NGC 4486B}$ is the central velocity dispersion of NGC 
4486B, are shown in Table 4 for the Na I, Ca I, $^{12}$CO (2,0), and $<FeI>$ indices; 
the impact of velocity smearing was not investigated for the $^{13}$CO index, since (1) 
the passbands were defined from spectra that were already velocity broadened, and (2) 
this index is not compared here with measurements made in other studies. Indices 
corresponding to a null velocity dispersion can be obtained by multiplying the values 
given in this study by the entries in Table 3. 
The C($\sigma$) entries in Table 4 are similar to those shown in Figure 8 of 
Silva et al. (2007) for a velocity dispersion of 330 km sec$^{-1}$.

\subsubsection{The Na I and Ca I Indices}

	The Na I and Ca I line indices sample the strengths of transitions 
from more than one element, and this complicates the 
sensitivities of the indices to chemical abundance. Such 
contamination is common among line indices at visible wavelengths (e.g. Worthey et 
al. 1994). In addition to Na lines, the Na I index also includes significant Si 
absorption, although Silva et al. (2007) point out that both Na and Si are expected to be 
super metal-rich in early-type galaxies, and this should at least partially mitigate the 
impact of contamination on the Na I index. Contamination is of greater potential concern 
for the Ca I index, where transitions due to S, Si, Ti, Sc, and Fe are present in 
addition to those of Ca. These elements have diverse chemical enrichment 
pedigrees, thereby complicating the interpretation of the Ca I index.

	Age is another parameter that might affect the indices. 
The depth of H$\beta$ is sensitive to the brightness of the main sequence turn-off, and 
galaxies that fall above the relations defined by old systems on the 
H$\beta$--[MgFe'] plane are thought to have a young stellar component. 
Silva et al. (2007) find that galaxies that fall well above the old galaxy relation 
on the H$\beta$--[MgFe'] plane also fall well away from the old sequence 
on the Na I--[MgFe'] diagram. Strong Na I absorption thus appears to be linked to strong 
H$\beta$ absorption, and so `aging' a system to predict the line indices 
that might be expected many Gyr in the future if there is no further star formation 
shifts the Na I index to smaller values. Silva et al. (2007) 
suggest that this behaviour is due to the presence of 
an extended giant branch, made up of AGB stars, in younger systems.

	The location of all three galaxies on the H$\beta$--[MgFe'] diagram were examined 
using the indices measured by Sanchez-Blazquez, Gorgas, Cardiel, \& Gonzalez (2006a) 
to determine if age should be considered when interpreting the near-infrared indices. 
NGC 4486B and NGC 5846A fall in a part of the diagram indicating that they are old 
systems, while M32 falls well above this relation. The Na I index measurements of 
M32 will thus have to be adjusted for age affects. In \S 5 it is argued that age may also 
affect the Na I indices in the central few tenths of an arcsec of NGC 4486B and NGC 5846A, 
which is a region that contributes only a modest fraction of the total signal 
to the Sanchez-Blazquez et al. (2006a) observations.

	The Na I and Ca I indices in each annulus are plotted against the 
$^{12}$CO(2,0) index in Figure 4. Ramirez et al. (1997) measured the strengths of Na I, 
Ca I, and $^{12}$CO (2,0) features in a sample of disk giants, and these data 
can be used to help interpret the galaxy indices. Caution should of course be 
exercised when comparing stellar and galaxy index measurements as galaxies are 
composite stellar systems, and their spectra show the integrated 
contributions from stars spanning a range of luminosities and properties. Still, given 
that the near-infrared spectral region is dominated by stars that are nearing the end of 
their evolution, the issue of dealing with integrated spectra is not as 
critical as at visible wavelengths, where the light comes from stars spanning a 
much more diverse range of evolutionary states. Indeed, in old systems 55\% of the 
$K-$band light from RGB stars, whereas RGB and main sequence stars 
each contribute 30\% of the $V-$band light (Maraston 2005).

	A Galactic disk sequence for the (Na I,$^{12}$CO) and (Ca I,$^{12}$CO) 
diagrams was obtained by fitting a linear relation to the entries in Table 5 of 
Ramirez et al. (1997) with $^{12}$CO between 10\AA\ and 16\AA, and the results 
are shown as dashed lines in the two top panels of Figure 4. 
To adjust for velocity smearing in the cE spectra, the Ramirez et al. (1997) 
indices were divided by the C$_{NGC4486B}$ values listed in Table 4 prior to 
fitting the linear relation. While the stars 
used by Ramirez et al. (1997) likely have a mix of abundances, the scatter 
about the mean relations is only on the order of $\pm 0.2\AA$. 

	The NGC 4486B and NGC 5846A data define a single $\sim 0.5\AA$ wide
sequence in the (Na I, $^{12}$CO) diagram, suggesting that there are similarities in the 
stellar contents of the central regions of these galaxies. More specifically, 
that NGC 4486B and NGC 5846A fall along a moderately tight sequence in the (Na I, 
$^{12}$CO) diagram suggests that the stars in both galaxies have similar Na $+$ Si 
abundances. It might be anticipated that the stars in elliptical galaxies experienced 
chemical enrichment histories that are systematically different from those of stars 
in the Galactic disk, with the result that stars in the galaxies 
may have chemical mixtures that differ from those in the Galactic disk. 
In fact, the Na I indices in NGC 4486B and NGC 5846A fall well above the Galactic disk 
sequence, by an amount that exceeds the $\pm 0.2\AA$ scatter in the Galactic disk 
giants.

	The M32 data are distinct from the NGC 4486B $+$ NGC 5846A 
sequence in the (Na I, $^{12}$CO) plane, in the sense of having weaker Na I absorption at 
a given $^{12}$CO(2,0) equivalent width. The M32 datapoints huddle in a very 
tight clump $\sim 0.4\AA$ above the Galactic disk giant sequence. 
The `aging vectors' on the H$\beta$--Na I plane in Figure 16 of Silva et al. 
(2007) suggest that as M32 ages then the Na I index will ultimately decrease by 0.4 -- 
0.5 \AA\ with respect to the present day value. Such a correction moves the M32 datapoints 
on the (Na I, $^{12}$CO) diagram in Figure 4 so that they fall directly on the trend 
defined by solar neighborhood giants. This suggests that the abundances of Na and Si in 
M32 are similar to solar neighborhood values.

	While there is considerable scatter, the galaxy datapoints on the (Ca I, 
$^{12}$CO) diagram overlap with the Galactic disk sequence. The M32 Ca I indices scatter 
immediately above the Galactic disk sequence in the middle panel of Figure 4, although the 
offset between the M32 and Galactic disk sequences is comparable to the scatter 
in the Galactic disk Ca I values. The locations of all three galaxies on the 
(Ca I, $^{12}$CO) diagram suggests that the elements that dominate the Ca I 
bandpass in these systems have abundances that are comparable to those among stars in the 
solar neighborhood. Thomas, Maraston, \& Bender (2003) conclude that the enrichment 
history of Ca in early-type galaxies may differ from that of other 
$\alpha-$elements, in that the Ca abundance is depressed with respect to other 
$\alpha-$elements. While caution should be excercised because of contamination 
from elements other than Na and Ca, that the NGC 4486B and NGC 5846A Ca I indices in 
Figure 4 scatter about the solar neighborhood sequence, whereas the 
Na I indices fall well above the solar neighborhood track in this figure, is 
consistent with the Thomas et al. (2003) findings.

\subsubsection{The $<FeI>$ Index}

	Silva et al. (2007) conclude that the $<FeI>$ index is dominated by Fe-peak 
elements. Still, CN absorption may compromise the metallicity sensitivity of this index, 
and could explain why it does not change amongst old galaxies with velocity 
dispersions larger than 150 km sec$^{-1}$ (Silva et al. 2007). Like the Na I 
index, the $<FeI>$ index is higher at a given [MgFe'] in galaxies that contain 
young populations than those that are believed to be old, although the offset between 
the young and old sequences is smaller than that shown by the Na I index.

	The NGC 4486B and NGC 5846A data form a tight sequence on the ($<FeI>$, 
$^{12}$CO) diagram, that falls $\sim 0.3\AA$ above the sequence defined by the solar 
neighborhood standard stars. The dispersion in the $<FeI>$ sequence is roughy consistent 
with the measurement errors. That the $<FeI>$ index does not change with $^{12}$CO in NGC 
4486B and NGC 5846A is consistent with the absence of a trend 
between $<FeI>$ and velocity dispersion when the latter is moderately high. 
The M32 data points fall near the lower end of the scatter 
envelope defined by NGC 4486B and NGC 5846A. The location of M32 on the 
(H$\beta$, [MgFe']) diagram suggests that the $<FeI>$ index of M32 will drop by 
$\sim 0.2\AA$ over the next few Gyr if there is no more star formation, and this 
is sufficient to move M32 onto the solar neighborhood sequence. 

\subsubsection{The $^{13}$CO(2,0) Index}

	The $^{13}$CO(2,0) index was measured in the spectra of the radial velocity 
standard stars, and the results were used to construct the Galactic disk 
reference sequence that is shown as a dotted line in the lower panel of Figure 
4. The M32 $^{13}$CO(2,0) measurements show little dispersion on the ($^{13}$CO,$^{12}$CO) 
diagram, and fall squarely on the standard star sequence. The NGC 4486B 
and NGC 5846A $^{13}$CO(2,0) measurements show a much larger dispersion than the 
M32 data, but still tend to scatter about the Galactic disk sequence.
$^{12}$C is produced in Type II SNe and in AGB stars, 
whereas $^{13}$C is produced in the CN cycle and ejected from 
intermediate mass AGB stars (Iben \& Renzini 1983). That the cE data points in the 
lower panel of Figure 4 scatter about the standard star sequence suggests that 
the enrichment timescale of $^{13}$C in these galaxies has been the same as in 
the solar neighborhood. 

\subsection{Absorption Line Gradients}

	The spatial distribution of stars within galaxies provides a fossil record of 
the galaxy's collapse history and subsequent evolution. 
The Na I, Ca I, $<FeI>$, and $^{12}$CO(2,0) indices are plotted as a function of 
radius in Figure 5. The errorbars show $1\sigma$ uncertainties 
that were computed from the dispersion in index measurements made from unstacked spectra. 
The uncertainties in each dataset are largest in the innermost annulus, where only 3 
individual spectra were combined. The horizontal axes in these figures are logarithmic 
with radius, as previous studies have found roughly linear relations between the 
equivalent widths of various absorption features and log(r) (e.g. Davidge 1992; Davies 
et al. 1993). Simulations also predict that evolutionary processes 
often produce linear relations between metallicity and log(r) (e.g. Kobayashi 2004).

	In an effort to quantify the size and significance of any gradients, least 
square fits were made to the indices as a function of log(r), and the resulting values of 
$\Delta$Index/$\Delta$log(r) are shown in Table 5. The plots in Figure 5
suggest that trends at larger radii may differ from those in the inner 
regions of the galaxies. Therefore, $\Delta$Index/$\Delta$log(r) was also computed 
only when $r > 0.5$ arcsec, and the results are shown in brackets in Table 5.

\subsubsection{M32}

	For the purposes of discussion, a gradient is considered to be significant 
when $\Delta$Index/$\Delta$log(r) differs from zero in excess of the $2\sigma$ level. 
In the case of M32, the gradient with the highest statistical significance is that 
involving the Ca I index, which decreases in strength with 
increasing radius when the full range of radial distances is considered. However, 
when only data with $r > 0.5$ arcsec are considered then a significant gradient is not 
found. While the $<FeI>$ index shows a significant gradient when 
$r > 0.5$ arcsec, the Na I and $^{12}$CO(2,0) indices do not show 
significant gradients. The low amplitudes or absence of gradients in the NIFS 
observations of M32 are consistent with line index measurements of this galaxy 
at visible wavelengths, which reveal line gradients that are weak or non-existent, and 
hence differ in character from what is typically seen in classical elliptical 
galaxies (e.g. Davidge 1991).

\subsubsection{NGC 4486B}

	The radial behaviours of the Ca I and $^{12}$CO(2,0) indices in 
NGC 4486B differ from what is seen in the other galaxies. There is a 
general tendency for the Ca I and $^{12}$CO(2,0) indices to strengthen with increasing 
radius near the galaxy center, with the trend either flattening or reversing direction 
near $r \sim 0.5$ arcsec. This change in behaviour affects the significance of gradients 
that are measured from the full range of radial distances covered by the NIFS data. 
For example, the entries in Table 5 indicate that the Ca I 
index increases with radius throughout all of the region that is covered by the NIFS data, 
although the gradient ceases to be significant if only the 
data with $r > 0.5$ arcsec are considered. For comparison, the $^{12}$CO(2,0) 
index in NGC 4486B shows a significant gradient in the sense of becoming 
weaker with increasing radius when $r > 0.5$ arcsec, although when the entire 
range of radii is considered then a significant gradient is found in the opposite sense, 
due to the behaviour of the $^{12}$CO(2,0) index near the nucleus. The 
Na I index does not vary significantly with radius in NGC 4486B.

	There is a significant gradient in the  $<FeI>$ index in NGC 4486B when the 
full range of radii is considered, while the slope differs from zero at the 
$1.9-\sigma$ level when only the measurements with $r > 0.5$ arcsec are considered. The 
measured slopes are skewed by the $<FeI>$ index at the largest radius, which is an outlier 
and is the smallest $<FeI>$ index measured in the three cEs. Still, even if this 
point is not considered then the slope measured from all of the remaining data 
differs from zero at more than the $2-\sigma$ level. Thus, a robust result is 
that there is a mild $<FeI>$ gradient in NGC 4486B.

	The complex radial behaviour of line indices in NGC 4486B is contrary to what 
might be expected from previous observations of this galaxy at visible wavelengths, 
which find that metallic absorption features systematically weaken with 
increasing radius (Davidge 1991; SGC). Indeed, the 
metallicity gradient in NGC 4486B measured by SGC is typical of 
that in classical elliptical galaxies. The NIFS observations of NGC 4486B have a much 
higher angular resolution than the data used by SGC and Davidge (1991), and 
these studies do not explore the nature of line gradients in 
the central 0.5 arcsec of this galaxy. The NIFS data indicate that the trends 
seen at large radius break down in the central arcsec of NGC 4486B.

\subsubsection{NGC 5846A}

	Both the Na I and Ca I indices vary significantly with radius in NGC 5846A, 
in the sense that they systematically weaken with increasing 
radius. The $^{12}$CO(2,0) index in NGC 5846A shows a similar tendency, although the slope 
differs from zero at only the $1.5-\sigma$ level. The $<FeI>$ index does not vary with 
radius in NGC 5846A. SGC found that metallic absorption features weaken with increasing 
radius in NGC 5846A, and the NIFS data suggest that the gradients that are present at 
larger radii extend into the central arcsec of this galaxy.

\subsubsection{NGC 4486B $+$ NGC 5846A: Two Peas in a Pod?}

	NGC 4486B and NGC 5846A show structural similarities, having similar overall 
sizes, in that the $r_{25}$ values in Table 1 correspond to 1.4 kpc for NGC 4486B and 1.6 
kpc for NGC 5846A, and central velocity dispersions. Thus, it is of interest to compare 
the line indices in the central regions of these galaxies. A complicating factor is that 
NGC 4486B and NGC 5846A are at different distances. To simulate the radial properties of 
NGC 4486B as if it were observed at the distance of NGC 5846A, the NGC 4486B measurements 
in Figure 5 were shifted by 0.19 dex to smaller radii, and the results are 
compared with the NGC 5846A measurements in Figure 6. {\it 
There is excellent overlap between the line indices in the two galaxies after accounting 
for the difference in distance.} When the two galaxies are considered in concert, there 
is a clear tendency for the Na I, Ca I, and $^{12}$CO(2,0) indices to weaken 
with increasing radius when log(r$_{NGC5846A}) > -0.5$. At smaller radii there is little 
or no evidence for radial variations in the Na I and Ca I indices. However, 
when log(r$_{NGC5846A}) < -0.5$ the $^{12}$CO(2,0) indices of 
both galaxies overlap, and show a clear tendency for the $^{12}$CO(2,0) index to 
weaken towards smaller radii. There is also a tendency for the $<FeI>$ 
indices to strengthen with decreasing radius when log($r_{NGC5846A}) < -0.9$. 
These comparisons suggest that the radial distributions of stellar content in the 
central regions of both galaxies have been influenced in similar ways. Possible causes 
of the line gradients in the innermost regions of these galaxies are considered in 
\S 5.

\section{DISCUSSION \& CONCLUSIONS}

	The near-infrared spectroscopic properties of the central arcsec of the 
cE galaxies NGC 4486B, NGC 5846A, and M32 have been investigated with 
NIFS $+$ ALTAIR on GN. As the closest cE galaxies, 
these are unique laboratories for probing the nature and origins 
of this galaxy type. The observations of NGC 4486B and NGC 5846A utilize 
the LGS capability on GN. The angular resolution is 
comparable to that obtained with the HST at visible wavelengths, and corresponds 
to a spatial resolution of $\sim 6$ parsecs at the distance of NGC 4486B and 
$\sim 12$ parsecs at the distance of NGC 5846A.

	Spectroscopic studies in the $K-$band are complementary to those at visible 
wavelengths. Whereas the integrated light at visible wavelengths comes from a 
diverse mix of stellar types, the $K-$band light from all but the youngest stellar 
systems is dominated by the most highly evolved first ascent giants and AGB 
stars, the photometric and spectroscopic properties of which are sensitive to 
age and metallicity. Indices that probe portions of the spectrum 
that contain, among other features, Na I, Ca I, and Fe I 
lines, as well as the first overtone bands of CO, have been measured in 
each galaxy. The relative strengths of these features and the manner with which they 
vary with radius can be used to compare the stellar contents of the 
central regions of the three galaxies, and provide insights into their past 
histories. The stellar content of each galaxy is discussed in turn below.

\subsection{M32}

	The NIFS data reveal weak or non-existent line gradients near the center 
of M32. Spectra of M32 at visible wavelengths show similar behaviour (e.g. Davidge 
1991), albeit at larger radii than probed with NIFS. Observations of M32 at 
visible wavelengths with the HST reveal that the $V-I$ color is constant 
throughout the central arcsec, although there may be a slight change in color within 
$\sim 0.05$ arcsec of the nucleus (Lauer et al. 1998). The absence of near-infrared line 
gradients near the center of M32 suggests that the uniform distribution of AGB stars 
that was found throughout the main body of M32 by Davidge \& Jensen (2007) continues 
into the innermost regions of the galaxy. 

	The first-overtone $^{12}$CO bands are sufficiently deep that their strengths 
can be gauged with narrow-band photometric measurements. Peletier (1993) investigated 
the radial profile of the CO index in M32, and the results in his Table 2 indicate 
that the CO index does not change with radius throughout much of the galaxy. 
However, within 2 arcsec of the galaxy center there is a tendency 
for the CO index to strengthen slightly with decreasing radius. 
While it should be kept in mind that the Peletier data likely have an angular resolution 
of an arcsec or more, the NIFS data also show a weak ($1.6\sigma$) tendency for 
$^{12}$CO(2,0) to strengthen with decreasing radius when $r > 0.5$ arcsec, although this 
trend disappears if data with $r < 0.5$ arcsec are included.

	The line indices measured from the NIFS spectra indicate that 
the chemical mixture in M32 differs from that in NGC 4486B and NGC 5846A. 
When placed on diagrams relating the Na I, Ca I, and $^{13}$CO(2,0) 
indices to the $^{12}$CO(2,0) index, M32 falls near or directly 
on the sequence defined by solar neighborhood giants, whereas 
NGC 4486B and NGC 5846A depart from the solar neighborhood giant trend on the 
(Na I, $^{12}$CO) diagram (\S 4). Given that the Na I and Ca I indices 
each sample more than one element, then these results suggest that 
the mixture of elements that contribute to the Na I and Ca I indices 
are close to solar in M32, as is the ratio $^{13}$C/$^{12}$C.

	The comparisons in Figure 4 thus suggest that the enrichment 
history for the gas from which the stars in M32 formed was similar 
to that in the Galactic disk, as opposed to the rapid enrichment that is thought to have 
prevailed in classical elliptical galaxies, and resulted in a relative deficiency in Fe 
(e.g. Trager et al. 2000). Evidence that M32 is not as Fe-deficient as other spheroidal 
systems is also seen at optical wavelengths, as M32 falls along the upper envelope 
of the Fe5270 and Fe5335 versus Mg$_2$ diagrams (e.g. Figure 2 of Worthey, Faber, \& 
Gonzalez 1992). In the context of the various formation models discussed in \S 1, 
a solar neighborhood-like chemical mixture is consistent with models in which 
stars in M32 formed from gas and dust that had been in a chemically mature spiral galaxy, 
in agreement with the `threshed' spiral model of Bekki et al. (2001). Models in which 
M32 is the end product of the stripping of a classical elliptical galaxy are ruled out.

\subsection{NGC 4486B}

	NGC 4486B has almost certainly had a highly eventful past. The galaxy is 
located in the core of the Virgo cluster, and so has been subjected 
to tidal interactions. Faber et al. (1997) note that the 
central photometric properties of NGC 4486B are such that 
it could be the remnant of an elliptical galaxy that lost $\sim 90\%$ of its mass through 
stripping. In fact, the location of NGC 4486B on the (Na I, $^{12}$CO) 
diagram is consistent with the majority of its stars having formed from gas and dust that 
experienced rapid chemical enrichment, as is the case in large spheroidal systems. 
Still, that the NGC 4486B datapoints scatter about the solar neighborhood 
giant sequence on the ($^{13}$CO,$^{12}$CO) diagram suggests that the enrichment could 
not have been so rapid as to preclude enrichment from AGB stars. In any event, 
the chemical enrichment history of NGC 4486B differs from that of M32. 

	While large-scale stripping of stars could explain the relatively high black 
hole mass computed for NGC 4486B by Kormendy et al. (1997), such catastrophic 
interactions with another galaxy would affect the spatial distribution of stars, and might 
be expected to obliterate population gradients that were in place 
in the progenitor (e.g. White 1978). It then becomes difficult to explain why 
the radial variation in mean metallicity in NGC 4486B is 
typical of what is seen in classical ellipticals (SGC). The situation 
might be different if a large reservoir of gas were present during 
the proposed interactions that stripped material from the progenitor. In this case 
the tidal interactions might funnel gas to the center of the 
progenitor, where star formation would be triggered. If dissipation occurs then 
metallicity gradients might be set up amongst the newly formed stars. While 
large reservoirs of cool gas are in short supply near the center of the Virgo cluster 
at the present day, the old age of NGC 4486B suggests that the structural characteristics 
of the galaxy could have been defined early-on in the life of the cluster, at an epoch 
when cool gas was more plentiful.

	Absorption line gradients, in the sense of weakening metallic 
features with increasing radius, are seen in the outer regions of NGC 4486B at 
visible wavelengths. However, neither the Na I nor the Ca I indices 
measured from the NIFS observations of NGC 4486B follow these trends. 
While there is a tendency for the $^{12}$CO(2,0) index to systematically weaken 
with increasing radius when $r > 0.5$ arcsec, within a few tenths of an arcsec 
of the center of NGC 4486B the $^{12}$CO index, like the Ca I index, 
{\it weakens} with decreasing radius. What mechanism can explain such a 
radial behaviour in the near-infrared line indices?

	The location of NGC 4486B on the 
H$\beta$--[MgFe'] diagram and the analysis by SGCN indicates 
that a young population does not dominate this galaxy on arcsec 
scales, and so the Na I and $<FeI>$ indices in the outer regions of the NIFS data are 
probably not affected by age. However, SGB find that there is an age gradient in NGC 
4486B with $\Delta$log(t)/$\Delta$log(r) $= 0.4 \pm 0.2$. This is one of the steepest age 
gradients in the SGC sample, and opens the possibility that there could 
be an intermediate age population near the center of NGC 4486B. 

	The radial behaviour of two of the near-infrared indices support the 
notion that the central regions of NGC 4486B harbour intermediate age stars. 
First, the $<FeI>$ index is a potentially powerful age indicator in galaxies that are 
dominated by metal-rich stars, as there is no metallicity sensitivity (Silva et al. 
2007). It is thus significant that this index shows a significant gradient in NGC 4486B, 
in the sense of getting stronger with decreasing radius. Second, the $^{12}$CO(2,0) 
index in the innermost regions of NGC 4486B are also smaller than at large radii, 
with an equivalent width that is similar to that near the center of M32. Models 
of simple stellar systems predict that the first overtone CO bands will grow in 
strength as a system ages when age exceeds 2 Gyr (e.g. Figure 11 of Vazquez 
et al. 2003).

	Based on the presence of a steep age gradient inferred from spectra at 
visible wavelengths, and the radial behaviour of the $<FeI>$ and $^{12}$CO indices, 
we suggest that the nuclear regions of NGC 4486B 
contain an intermediate age population. This possibility can be tested using 
space-based observations, where high angular resolution 
observations can be obtained at visible and blue wavelengths. If a young or intermediate 
age population is present in the nuclear regions of NGC 4486B then such data 
should show deeper Balmer absorption lines than at large radii. 
The line indices measured by Sanchez-Blazquez et al. (2006a) indicate that 
H$\beta$ near the center of NGC 4486B should be $\sim 0.5\AA$ higher than 
in the outer regions if the center of NGC 4486B harbours a population 
with an age like that in M32 (i.e. near $\sim 3$ Gyr).
There is a high frequency of distinct nuclei in Virgo 
Cluster early-type galaxies (e.g. Cote et al. 2006), and the colors of some 
nuclei are suggestive of an intermediate age population. NGC 4486B may 
thus be yet another of these systems showing a distinct nuclear population. 

	The notion that age drives the gradients in the central regions 
of NGC 4486B is not wihout its problems. Indeed, Lauer et al. (1996) 
find that the $V-I$ color of NGC 4486B tends to become {\it redder} 
towards smaller radii, and this is contrary to what would be expected if there 
were a moderately bright population of blue main sequence turn-off stars 
near the center of the galaxy. Therefore, it is worth considering 
other mechanisms that might create spectroscopic gradients near the center of 
NGC 4486B. 

	Centrally-concentrated continuum emission can veil absorption features, and cause 
line index trends that are defined at large radii to break down near galaxy centers. An 
excellent example of this is seen in the spectrum of M87, where the strengths 
of absorption features decrease with decreasing radius near the galaxy center 
due to LINER continuum emission (e.g. Davidge 1992). 
However, the central regions of NGC 4486B are devoid of nuclear activity. 
The nuclear x-ray emission from NGC 4486B is a point 
source when imaged with Chandra (Soria et al. 2006b), and the x-ray spectrum 
is softer than that expected from AGN. The spatial extent of the 
x-ray emission is $<< 70$ parsecs, and so there is probably 
little or no diffuse hot gas near the center of the galaxy. 
The visible spectra are free of emission lines, and there is no 
evidence of $2.12\mu$m H$_2$ emission in the NIFS data, which is a prominent 
feature in AGN spectra.

	The presence of a supermassive black hole will have an impact on population 
gradients in the inner regions of galaxies. Adopting a central black hole mass of 
$6_{-2}^{+3} \times 10^8$ M$_{\odot}$ and a velocity dispersion of 116 km sec$^{-1}$ 
(Kormendy et al. 1997) then the sphere of influence (van der Marel 1999) of the 
black hole in NGC 4486B has a radius $\sim 1.9_{-0.6}^{+1.0} \times 10^2$ parsecs, or 
$\sim 2.5_{-0.7}^{+1.2}$ arcsec. The gravitational field of the black hole dominates 
stellar motions within this radius, and any population gradients that date 
from the initial formation of the galaxy will be altered. This will occur on 
a timescale of only a few crossing times of the central regions, or $\sim 10^7$ years. 
The sphere of influence computed with this black hole mass is sufficiently large that 
some impact on the radial behaviour of line indices should have been seen 
in visible wavelength spectra recorded with arcsec 
resolution, and evidence for this has not been detected.

	The size of the sphere of influence depends on the mass of 
the central black hole. Soria et al. (2006a) 
compute a black hole mass of only 5 $\times 10^7$ M$_{\odot}$ for NGC 4486B using an 
empirical relation between light profile shape and black hole mass, and note 
that other empirical relations predict similar masses. 
While the black hole mass estimated by Kormendy et al. (1997) would produce 
a sphere of influence with a radius that exceeds the region studied with NIFS, 
if the Soria et al. (2006a) mass estimate is adopted then the sphere of influence 
has a $\sim 0.3$ arcsec radius. This is comparable to the angular scale where 
a break is seen in the behaviour of the CO and Ca I indices in NGC 4486B, and 
where the structural properties of the galaxy are observed to change (Lauer et al. 
1996). Still, a model in which the SMBH has altered population gradients in NGC 
4486B is not without its problems. For example, it might be anticipated 
that the population trends defined at large radii would flatten -- but 
not necessarily reverse -- near the galaxy center.

\subsection{NGC 5846A}

	The NGC 5846A and NGC 4486B datapoints occupy similar regions on the 
(Na I, $^{12}$CO), (Ca I, $^{12}$CO), and ($^{13}$CO,$^{12}$CO) diagrams, 
and this suggests that their chemical enrichment pedigrees -- and hence formation 
timescales -- were similar. In addition, while the purpose of this study is not 
to probe the kinematic properties of these galaxies, the central velocity dispersion 
measured in NGC 5846A is comparable to that in NGC 4486B. 
This suggests that the two galaxies may harbour central 
black holes with comparable masses. 

	SGC investigated the strengths of various absorption features 
in the visible spectrum of NGC 5846A, and found that metallic absorption 
features systematically weaken with increasing radius. Statistically 
significant gradients are seen at large radius in the Na I and Ca I indices 
measured from the NIFS spectrum of NGC 5846A. There is also weak, but not 
statistically significant, evidence for a radial gradient in the 
$^{12}$CO(2,0) index in the NIFS data. These results suggest that the 
gradients seen at larger radii extend into the central arcsec of NGC 5846A. 

	SGCN computed an old age for NGC 5846A. When coupled with the 
extremely old age measured for the dominant galaxy in the system -- NGC 5846  
-- by Trager et al. (2000), then it appears that large quantities of cold gas 
have not been present in the NGC 5846 $+$ NGC 5846A 
system for a large fraction of the age of the Universe.
Still, while the bulk of NGC 5846A has an old age, SGN measure an 
age gradient $\Delta$log(t)/$\Delta$log(r) $= 0.2 \pm 0.1$, and so 
the central regions might harbour stars that are younger than those at large radii. 
The comparison in Figure 6 suggests that the radial distribution 
of stellar content in NGC 5846A is similar to that in NGC 4486B, 
where the near-infrared indices suggest that a central intermediate age 
population is present. As with NGC 4486B, it would be interesting to obtain high angular 
resolution spectra of NGC 5846A at visible wavelengths to search for 
evidence of a nuclear enhancement in Balmer absorption line strengths, 
although the larger distance of NGC 5846A will make it more challenging to 
detect a nucleated young population than in NGC 4486B. 

\acknowledgements{It is a pleasure to thank an anoymous referee for a 
comprehensive report that lead to significant improvements in the manuscript.}

\clearpage

\begin{table*}
\begin{center}
\begin{tabular}{lcccc}
\tableline\tableline
Galaxy & $\mu_0$ & M$_K$ & r$_{25}$ & $\sigma_{0.1}$ \\
 & & & (arcsec) & (km sec$^{-1}$) \\
\tableline
M32 & 24.55 & --19.5 & 261.3 & 140 \\
NGC 4486B & 31.03 & --20.9 & 18.1 & 330 \\
NGC 5846A & 31.98 & --22.1 & 13.1 & 310 \\
\tableline
\end{tabular}
\end{center}
\caption{Seleted Properties of the cE Galaxies}
\end{table*}

\clearpage

\begin{table*}
\begin{center}
\begin{tabular}{lll}
\tableline\tableline
Star & Spectral Type & Standard Type \\
\tableline
HR4973 & K1III & Spectroscopic \\
HR4801 & K5III & Spectroscopic \\
HR4884 & M0III & Spectroscopic \\
 & & \\
HD13869 & A0V & Telluric (M32) \\
HIP116449 & A0V & Telluric (M32) \\
FS-131 & F8 & Telluric (NGC 4486B) \\
HD129153 & F0V & Telluric (NGC 5846A) \\
HD145228 & F0V & Telluric (NGC 5846A) \\
\tableline
\end{tabular}
\end{center}
\caption{Telluric and Spectroscopic Standard Stars}
\end{table*}

\clearpage

\begin{table*}
\begin{center}
\begin{tabular}{lc}
\tableline\tableline
Position & Wavelength Range \\
 & (\AA) \\
\tableline
Continuum & 23408 -- 23418 \\
$^{13}$CO(2,0) & 23418 -- 23476 \\
Continuum & 23476 -- 23486 \\
\tableline
\end{tabular}
\end{center}
\caption{Passbands for the $^{13}$CO(2,0) Index}
\end{table*}

\clearpage

\begin{table*}
\begin{center}
\begin{tabular}{ccccc}
\tableline\tableline
Spectral & C$_{Na I}$ & C$_{Ca I}$ & C$_{^{12}CO(2,0)}$ & C$_{<FeI>}$ \\
Type & & & & \\
\tableline
M0 III & 1.28 & 1.05 & 1.10 & 1.19 \\
K1 III & 1.34 & 1.10 & 1.12 & 1.27 \\
\tableline
\end{tabular}
\end{center}
\caption{C($\sigma_{NGC4486B}$) Values}
\end{table*}

\clearpage

\begin{table*}
\begin{center}
\begin{tabular}{lccc}
\tableline\tableline
Index & NGC 4486B & NGC 5846A & M32 \\
\tableline
Na I & $-0.14 \pm 0.10$ & $-0.53 \pm 0.21$ & $0.04 \pm 0.05$ \\
 & ($-0.97 \pm 0.44$) & ($-1.36 \pm 0.91$) & ($0.00 \pm 0.21$) \\
 & & & \\
Ca I & $0.65 \pm 0.20$ & $-0.82 \pm 0.31$ & $-0.18 \pm 0.04$ \\
 & ($2.95 \pm 1.04$) & ($-2.79 \pm 1.64$) & ($-0.07 \pm 0.20$) \\
 & & & \\
$<FeI>$ & $-0.20 \pm 0.09$ & $-0.09 \pm 0.10$ & $-0.10 \pm 0.06$ \\
 & ($-1.02 \pm 0.55$) & ($0.18 \pm 0.46$) & ($-0.27 \pm 0.11$) \\
 & & & \\
$^{12}$CO(2,0) & $-0.19 \pm 0.67$ & $-0.65 \pm 0.42$ & $0.03 \pm 0.08$ \\
 & ($-9.05 \pm 1.67$) & ($-3.67 \pm 1.90$) & ($-0.48 \pm 0.29$) \\
 & & & \\
\tableline
\end{tabular}
\end{center}
\caption{Line Gradients}
\end{table*}

\clearpage

\begin{figure}
\figurenum{1}
\epsscale{0.75}
\plotone{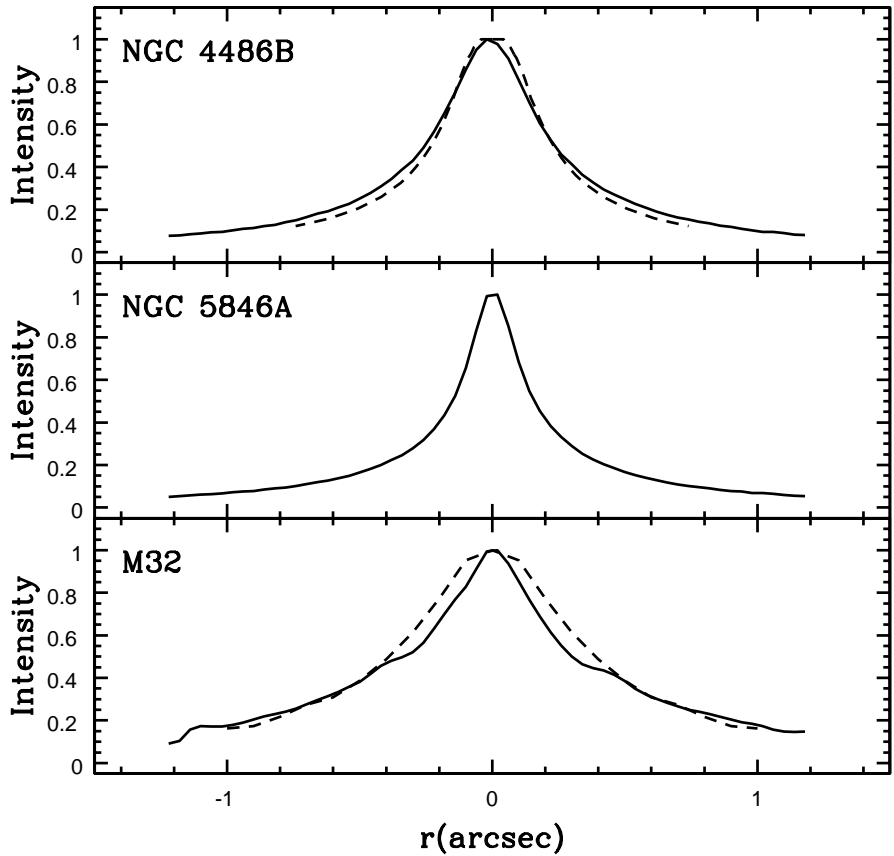}
\caption{The near-infrared light profiles of the three cE galaxies, as constructed 
from the NIFS observations. The solid line in each panel shows the 
light profiles obtained by integrating the two-dimensional spectra between 
rest frame wavelengths 2.2 and $2.27\mu$m, while the dashed lines show light 
profiles from HST observations. All of the profiles have been normalised to their peak 
values to facilitate comparisons. The HST profile shown for NGC 4486B is from Ferrarese et 
al. (2006), and there is excellent agreement with the profile constructed from 
the NIFS data. The HST data shown in the lower panel is the F222M 
profile from Figure 8 of Corbin et al. (2001). That the HST light profile near the 
center of M32 is broader than that observed with NIFS suggests that the present data 
have a higher angular resolution that that delivered by NICMOS.}
\end{figure}

\clearpage

\begin{figure}
\figurenum{2}
\epsscale{0.75}
\plotone{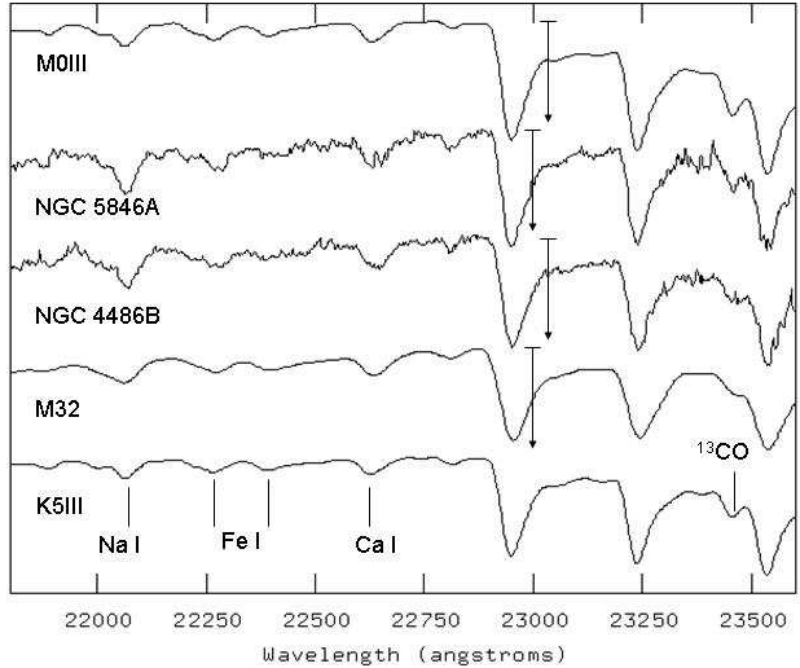}
\caption{The integrated spectra within 1 arcsec radius of the centers 
of M32, NGC 4486B, and NGC 5846A. Spectra of the solar neighborhood stars 
HR 4801 (K5 III) and HR 4884 (M0 III) are also shown. The spectra have been 
smoothed to a common velocity dispersion using the procedure described in the text, and 
the results were divided by a low-order continuum. The horizontal axis shows rest 
frame wavelengths, while the vertical axis is a linear intensity scale; the spectra have 
been shifted vertically to facilitate comparison. The locations of absorption 
features due to Na I, Fe I, Ca I, and $^{13}$CO(2,0) 
are indicated, while the (2,0), (3,1), and (4,2) $^{12}$CO band heads are the 
deep features at the red end of the spectra. The vertical arrows near 23000\AA\ 
show the depths of the $^{12}$CO (2,0) band head in the K5III spectrum. Note the 
galaxy-to-galaxy differences in the depths of the $^{12}$CO band heads.}
\end{figure}

\clearpage

\begin{figure}
\figurenum{3}
\epsscale{0.75}
\plotone{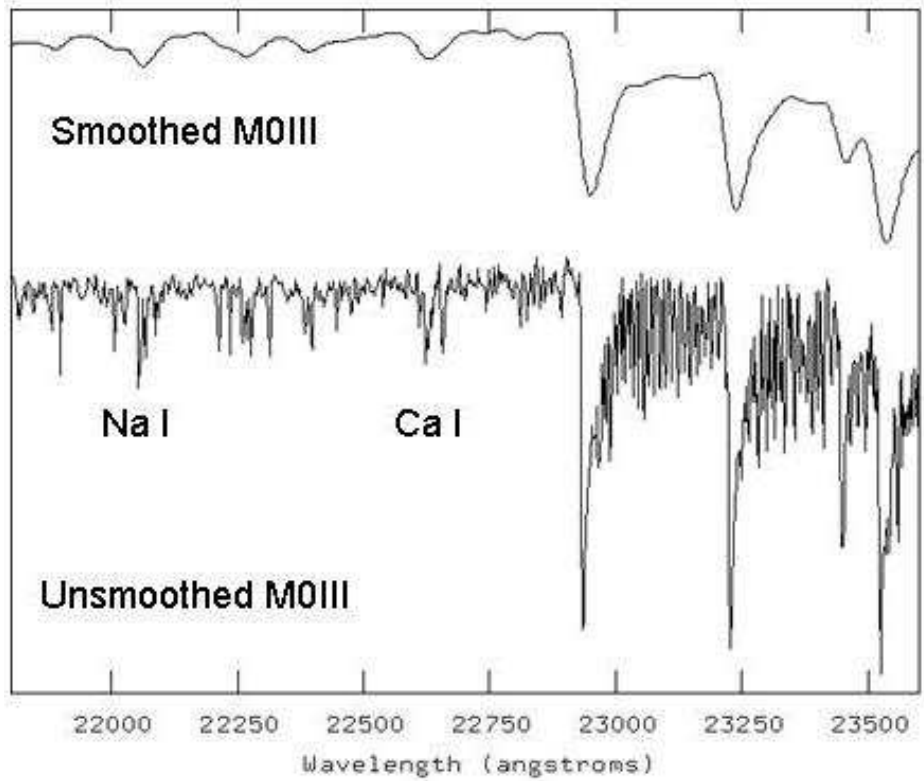}
\caption{The impact of kinematic broadening on the $2\mu$m spectrum of the M0III giant HR 
4884. The top spectrum shows the result of convolving the lower spectrum with a Gaussian 
to simulate the effects of kinematic smearing in the central regions 
of NGC 4486B. Note that the line complexes in the vicinity of the Na I and Ca I 
transitions appear as single, broad troughs in the smoothed spectrum.}
\end{figure}

\clearpage

\begin{figure}
\figurenum{4}
\epsscale{0.75}
\plotone{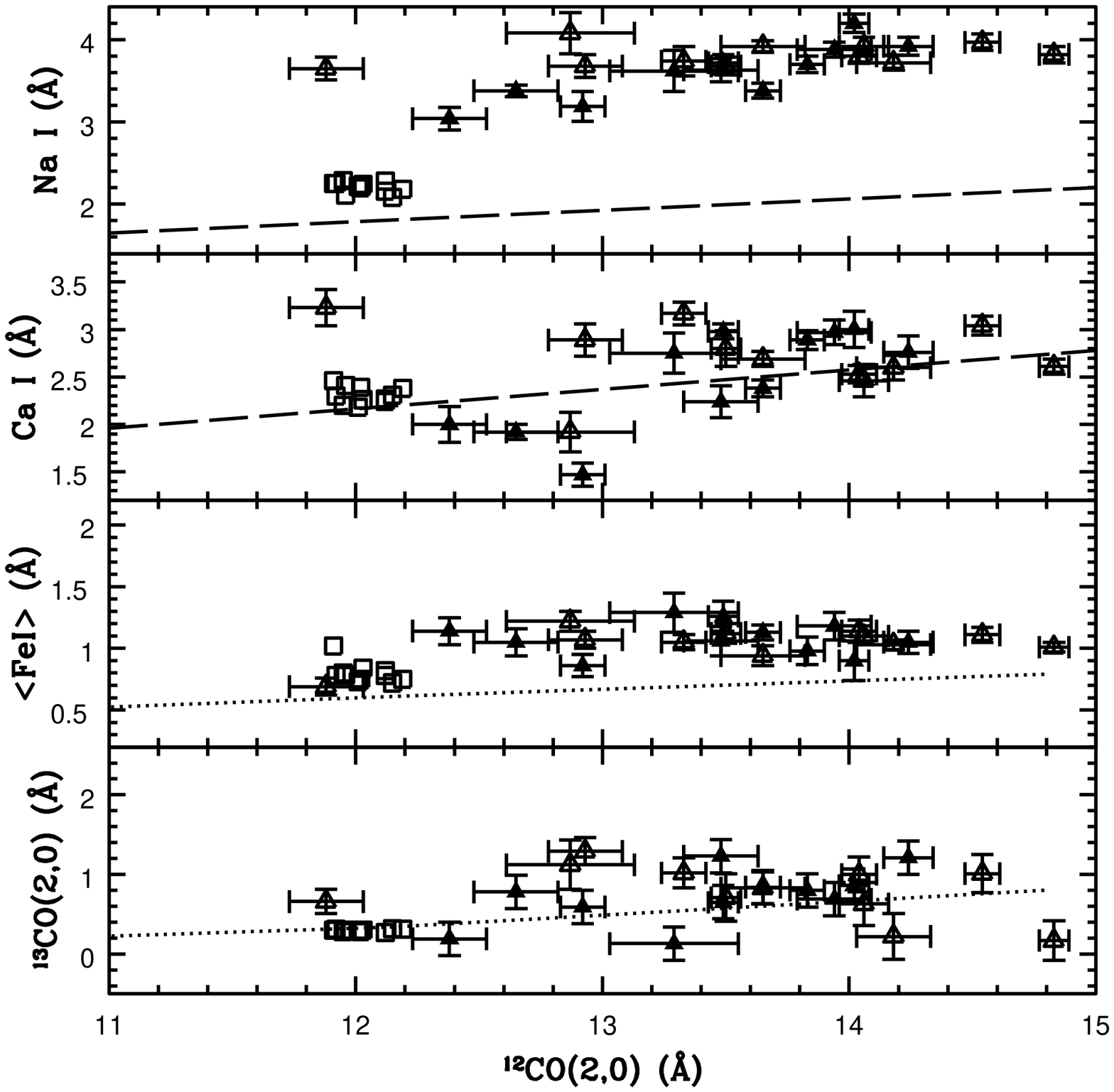}
\caption{The Na I, Ca I, $<FeI>$, and $^{13}$CO(2,0) indices in the spectra of 
NGC 4486B (open triangles), NGC 5846A (filled triangles), and M32 (open squares) 
as functions of the $^{12}$CO(2,0) index. The dashed line in the top two 
panels is the sequence defined by solar neightborhood giants with 
$^{12}$CO(2,0) between 10\AA\ and 16\AA\ from Table 5 of Ramirez et al. (1997), 
shifted to simulate the impact of velocity smearing. 
The dotted line in the lower two panels shows the sequence defined by giants in 
the solar neighborhood that were observed as radial velocity/spectroscopic standards 
with NIFS.} 
\end{figure}

\clearpage

\begin{figure}
\figurenum{5}
\epsscale{0.75}
\plotone{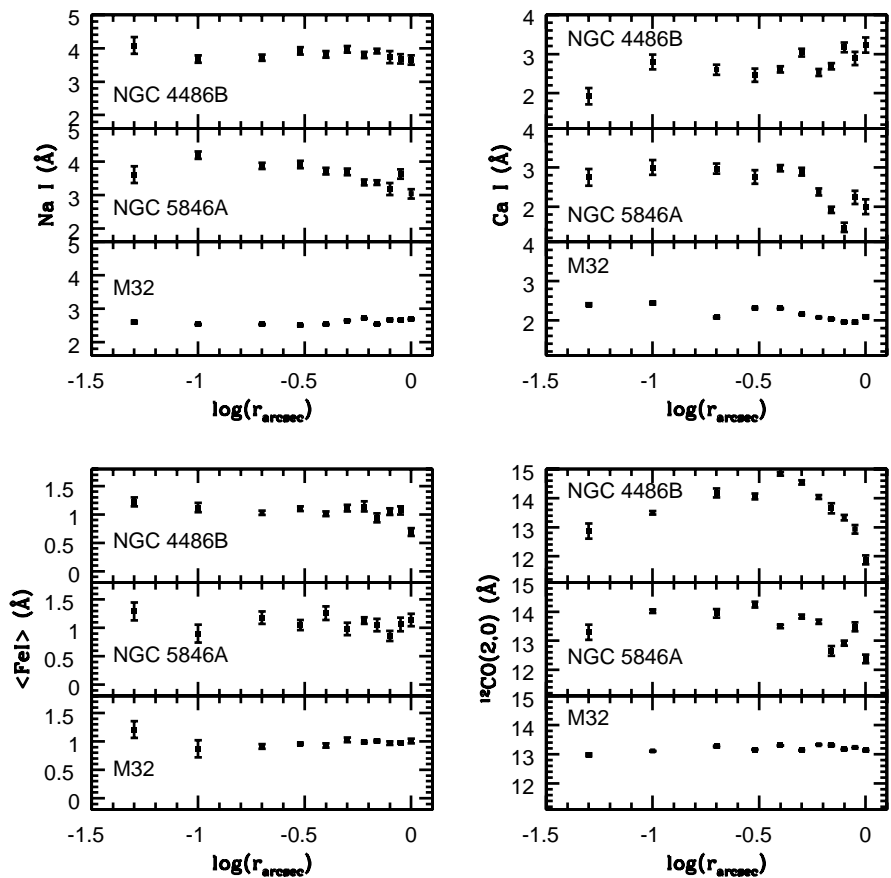}
\caption{The behaviour of near-infrared indices as a function of radius. 
The error bars show $1\sigma$ uncertainties computed from the scatter in indices 
measured in unstacked spectra. The uncertainties are largest in the central bin because of 
the small number of spectra that were combined there.}
\end{figure}

\clearpage

\begin{figure}
\figurenum{6}
\epsscale{0.75}
\plotone{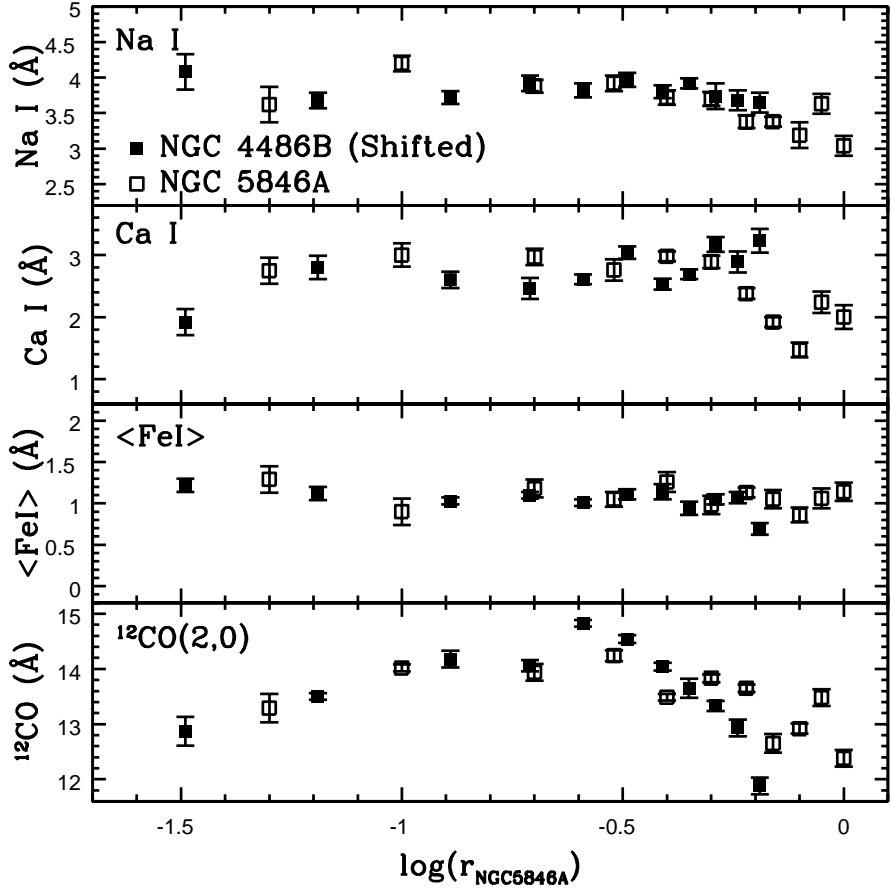}
\caption{The behaviour of the Na I, Ca I, and $^{12}$CO(2,0) indices with radius in NGC 
4486B (filled squares) and NGC 5846A (open squares) is investigated in this figure. 
The NGC 4486B measurements have been shifted towards smaller radii by 0.19 dex to 
simulate the radial behaviour of line indices if this galaxy were viewed at 
the same distance as NGC 5846A. r$_{NGC5846A}$ is the distance in arcsec 
at the distance of NGC 5846A. Note the excellent overlap between 
the two sets of indices, indicating that the radial distribution of stellar content 
in both galaxies has been shaped by similar physical processes.}
\end{figure}

\end{document}